\documentclass[%
 reprint,
superscriptaddress,
 amsmath,amssymb,
 aip,
 jcp, 
floatfix,
]{revtex4-2}

\usepackage{graphicx}
\usepackage{dcolumn}
\usepackage{bm}
\usepackage{xcolor}
\usepackage{soul}
\usepackage{hyperref}

\begin{document}

\title{High-dimensional order parameters and neural network classifiers applied to amorphous ices}

\author{Zoé Faure Beaulieu}
\author{Volker L. Deringer}
\email{volker.deringer@chem.ox.ac.uk}
\affiliation{Inorganic Chemistry Laboratory, Department of Chemistry, University of Oxford, Oxford OX1 3QR, United Kingdom}
\author{Fausto Martelli}
\email{fausto.martelli@ibm.com}
\affiliation{IBM Research Europe, Hartree Centre, Daresbury WA4 4AD, United Kingdom}
\affiliation{Department of Chemical Engineering, University of Manchester, Manchester M13 9PL, United Kingdom}

\begin{abstract}
Amorphous ice phases are key constituents of water's complex structural landscape. This study investigates the polyamorphic nature of water, focusing on the complexities within low-density amorphous ice (LDA), high-density amorphous ice (HDA), and the recently discovered medium-density amorphous ice (MDA). We use rotationally-invariant, high-dimensional order parameters to capture a wide spectrum of local symmetries for the characterisation of local oxygen environments. We train a neural network (NN) to classify these local environments, and investigate the distinctiveness of MDA within the structural landscape of amorphous ice. Our results highlight the difficulty in accurately differentiating MDA from LDA due to structural similarities. Beyond water, our methodology can be applied to investigate the structural properties and phases of disordered materials. 

\end{abstract}

\maketitle

Water is ubiquitous in everyday life and yet, due to its numerous anomalous behaviors, a complete understanding of its properties remains elusive. The complexity of water is reflected in the intricacy of its phase diagram, the most complex of any pure substance~\cite{Salzmann-19-2}. Since the discovery of polymorphism in crystalline ice by Bridgman in 1912 \cite{Bridgman-12}, over 15 new crystalline phases have been reported \cite{Salzmann-19-2}, with 3 discovered in the last 5 years \cite{Millot-19-5,Yamane-21-2,Prakapenka-21-11}.

At deeply super-cooled conditions water also possesses polyamorphism (amorphous polymorphism), a signature (but not a proof) of the existence of a liquid--liquid critical point~\cite{tanaka2020liquid}. Low-density amorphous (LDA) and high-density amorphous (HDA) ices \cite{Amann-Winkel-16-2, Loerting-11} are two classes of amorphous ices that encompass a broader set of sub-families characterized by differing structures, densities, and/or preparation routes. LDA is believed to be the most abundant form of ice in the Universe, formed from the cooling of water droplets onto the cold surfaces of dust in the interstellar medium. LDA can also be prepared via rapid quench of liquid water at ambient conditions, and its family comprises LDA-I and LDA-II, obtained via heating HDA and vHDA respectively \cite{Winkel-09-5}, as well as a more-ordered low-density phase obtained upon heating ice VIII \cite{Shephard-16-5}. HDA is formed via a first-order phase transition from compression of LDA or hexagonal ice I\textit{h}, resulting in a $\sim$20\% density increase \cite{Mishima-84-8}. Two other phases fall under the HDA bracket, namely, annealed and very high-density amorphous ice, eHDA \cite{Nelmes-06-6} and vHDA \cite{Loerting-01}. From a computational point of view, the differences between the sub-families of LDA and HDA are too small to be easily distinguished from their parents and, therefore, we will only refer to LDA and HDA. 

Amorphous ices are very complex. For instance, the short-range order in LDA and HDA is remarkably different. LDA shows high tetrahedrality resulting from well-separated first and second hydration shells. HDA, in contrast, is more disordered, with water molecules populating the space between the first and the second shells of neighbors and organized in motifs reminiscent of ice IV~\cite{Martelli-18-7,Shephard-17-4,Kobayashi-23-11}. LDA and HDA are also remarkably different in terms of the topology of the hydrogen bond network~\cite{Formanek-23-5}, but have comparable degrees of suppression of long-range density fluctuations~\cite{Martelli-17-9,Formanek-23-5}. 

The polyamorphism of water is of great interest for several reasons. For one, LDA and HDA are supposedly the glassy states of the low-density liquid (LDL) and the high-density liquid (HDL), which may be separated by a liquid--liquid phase transition at deeply under-cooled conditions. The existence of a liquid--liquid critical point has been proven by computer simulations for several models of water~\cite{Palmer-14-6,Debenedetti-20-7}, but the definitive experimental proof is still lacking (although experiments strongly support the hypothesis~\cite{Sellberg-14-6,Kim-20-11}). A division line of the first-order between LDA and HDA, as found both experimentally~\cite{Mishima-85-3} and from computer simulations~\cite{Giovambattista-05-9,Engstler-17-8,Formanek-23-5}, might extend at higher temperatures and end at the liquid--liquid critical point~\cite{Mishima-98-11}. Besides canonical hysteresis cycles, LDA and HDA embed signatures of metastable criticality in their long-range order~\cite{Martelli-17-9,Gartner-21-6,Formanek-23-5}.

Recently, a joint experimental and computational investigation reported the existence of another amorphous phase with a density in between that of LDA and HDA. This new amorphous ice, named medium-density amorphous ice (MDA), can be obtained by ball milling hexagonal ice at low temperatures~\cite{Rosu-Finsen-23-2}, a process that can be simulated via random shearing of layers of hexagonal ice~\cite{Rosu-Finsen-23-2}. The authors suggest various possible explanations, including MDA being the true glass of liquid water. Were this to be the case, it would question the validity of the liquid--liquid critical point hypothesis. To align with the two-state model, MDA would need to have a glass transition above the temperature of the liquid--liquid critical point, and thus would represent a metastable state before phase separation into LDA and HDA at the temperature of the liquid--liquid critical point~\cite{Rosu-Finsen-23-2}.

In the present study, we investigate the structural fingerprints of LDA, HDA, and MDA using Steinhardt bond orientational order parameters~\cite{Steinhardt-83-7} (BOOs). We exploit the fact that BOOs provide a representation of the rotation group SO(3), relating the irreducible representation of SO(3) and the symmetries of crystalline structure, hence allowing for a systematic investigation of crystalline symmetries. We follow Ref.~\citenum{Martelli-20-9} and encode all possible crystalline symmetries in a 30-dimensional vector containing the following OPs: $q_{l}(i)$ and $\bar{q}_{l}(i)$ with $l \in [3,12]$, $w_{l}(i)$ and $\bar{w}_{l}(i)$ with $l$ even and $l \in [4,12]$ (see Section 1.1. of the SI for further details). From these, we are able to gain an understanding of the structural properties of amorphous ices and elucidate the differences between LDA, HDA, and MDA. These BOOs are completely general and applicable to all disordered materials.

\begin{figure}[t]
    \centering
    \includegraphics[width=\linewidth]{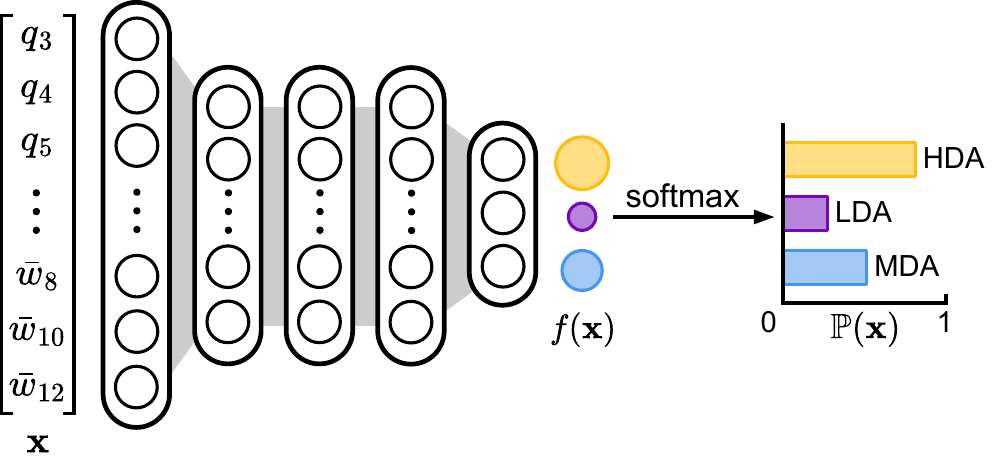}
    \caption{Schematic representation of the neural network employed in the present study. The input layer (\textbf{x}) consists of 30 nodes, each representing one of the structural bond orientational order parameters. The output layer has three nodes for the HDA, LDA, and MDA phases respectively. A softmax activation function is used on the output layer to convert the raw outputs, $f$(\textbf{x}), to predicted probabilities, $\mathbb{P}$(\textbf{x}).}
    \label{fig:nn_schematic}
\end{figure}

\begin{table}[t]
\setlength{\tabcolsep}{8pt}
\renewcommand{\arraystretch}{1.2}
\caption{Optimized NN hyperparameters.}
\begin{tabular}{lcc}
\hline
\hline
\multicolumn{1}{c}{\textbf{Hyperparameter}} & \textbf{Range}                      & \textbf{Optimized Value} \\ 
\hline
\# hidden layers                              & [1, 5]                              & 3                        \\
\# neurons per layer                                    & [8, 256]                          & 82                       \\
Weight decay                                 & [$10^{-8}$, 0.1] & $1.36\times10^{-4}$                 \\
Learning rate                                 & [$10^{-5}$, 0.1] & $6.36\times10^{-3}$                 \\ 
\hline
\hline
\end{tabular}
\label{table:NN_optimisation}
\end{table}

We consider the classification task to identify the class of structure (HDA, LDA, or MDA) from which a given atomic environment has been sampled. This approach builds upon  prior work by Martelli et al. \cite{Martelli-20-9}, wherein an NN was trained in a similar manner on HDA, LDA, and liquid structures. Herein, we train on pairs $(\mathbf{x}_i, y_i)$, where $\{\mathbf{x}_i\}$ are the 30-dimensional BOOs describing the local environment of atom $i$, and $\{y_i\}$ are one-hot encoded vectors relating the environment to exactly one of the 3 classes.

To learn this mapping, we adopt the simplest neural network (NN) architecture: a feed-forward multi-layer perceptron (MLP). We used ReLU activation functions \cite{agarap2019deep} for the hidden layers and a softmax activation function for the final layer. The latter transforms raw output values into predicted probabilities over the input classes (Fig.~\ref{fig:nn_schematic}). To optimize the weights of the network, we minimized the cross-entropy loss using the Adam optimizer \cite{kingma2017adam} as implemented in \texttt{PyTorch} \cite{paszke2017automatic}. We optimized the number of hidden layers, number of neurons per layer, learning rate and weight decay using Bayesian optimisation, as implemented in \texttt{Optuna} \cite{akiba2019optuna}. The sampled ranges and optimisation outcomes are found in Table \ref{table:NN_optimisation}. 

Interestingly, we observe that such a model is very insensitive to hyperparameter selection. Among 110 optimisation iterations, 97\% of models achieved test set accuracy within 3\% of the best model's performance. Our final model comprises 3 hidden layers, each containing 82 neurons. 

\begin{figure}[t]
\includegraphics[width=\linewidth]{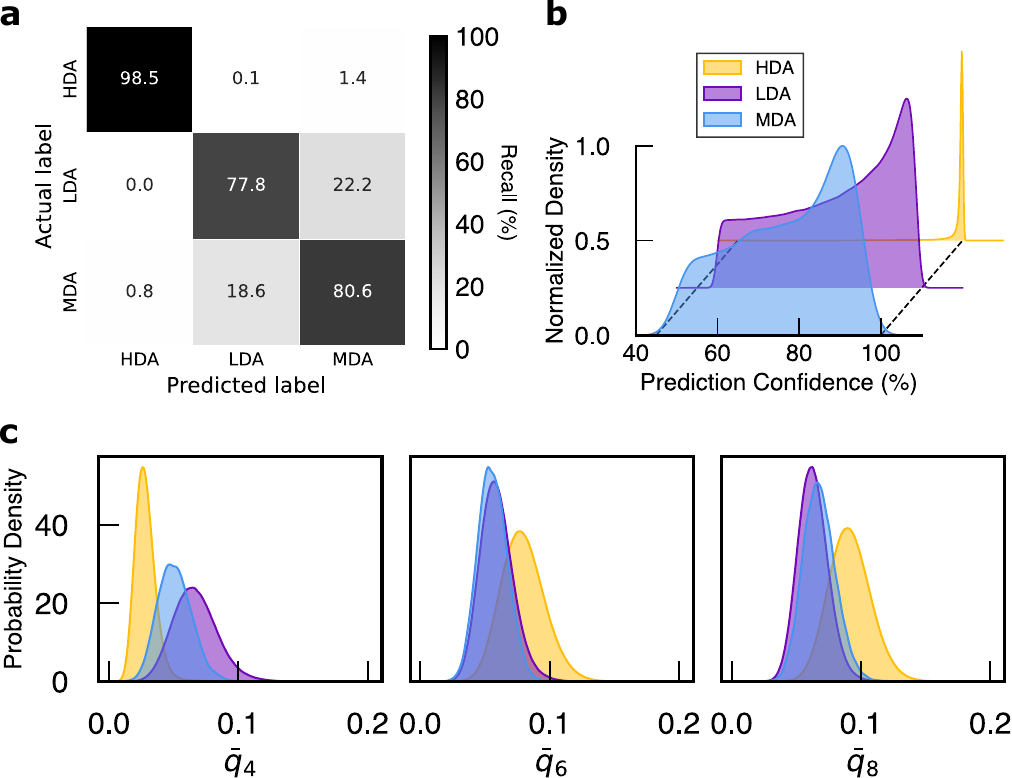}
\caption{Analysis of amorphous ices. (a) Confusion matrix for local environments classification in a test set containing HDA, LDA, and MDA. (b) Prediction confidence for the three target states. (c) KDE plots of BOOs $\bar{q}_{4}$, $\bar{q}_{6}$, and  $\bar{q}_{8}$ for the test set.}
\label{fig:mda_analysis}
\end{figure}

We use a confusion matrix to quantify the accuracy of our classification model (Fig.~\ref{fig:mda_analysis}a). The model shows very good performance in identifying HDA-like atoms but is notably worse at differentiating the two other target states, LDA and MDA. Martelli et al. \cite{Martelli-20-9} demonstrated that a similar NN trained exclusively on HDA, LDA, and liquid water achieved a  misclassification ratio of under 5\% between these three phases. These results are discussed in Section 2.1 of the SI. In contrast, the misclassification ratio between MDA and LDA increases to $\sim18-23\%$ upon the inclusion of MDA in the training dataset. This result suggests that LDA and MDA are quite hard to distinguish, at least from a structural point of view.
This is further demonstrated by Fig.~\ref{fig:mda_analysis}b which shows the confidence with which the model is classifying each target state. We observe a stark contrast between the model's confidence in classifying HDA environments compared to LDA and MDA environments. The much broader distribution of confidences demonstrates that the model is unable to accurately differentiate between LDA and MDA environments based on local order parameters alone.

The model's performance can be rationalized by inspecting the distribution of 3 BOOs for the three target states (Fig.~\ref{fig:mda_analysis}c). Excluding MDA, the distributions of HDA and LDA are distinctly separable, facilitating the NN's ability to categorize local environments based on their BOO values. However, with MDA included, this distinction becomes blurred. There is a substantial overlap between the LDA and MDA distributions indicating high structural similarities. Consequently, clear delineation among these phases in the structural space becomes ambiguous. This hampers the network's capability and confidence in classifying local environments which results in a greater rate of misclassification.

We benchmark our model's performance against various established classification algorithms, as implemented in \texttt{scikit-learn} \cite{scikit-learn}. Table \ref{table:bechmarking_table} shows the results of this analysis for the three most common classification models: k-Nearest Neighbors (k-NN), Logistic Regression, and Random Forest.  All three baseline models exhibit similar performance to our NN in classifying HDA, achieving recall rates of $\sim$98\%. As expected, when dealing with the more challenging task of differentiating LDA and MDA environments, all models struggle to achieve recall rates surpassing $80\%$, with k-NN notably underperforming in this regard. The balanced accuracy scores (BAS), which summarize  overall model accuracy, show that our NN outperforms the benchmark models for this classification task. For additional details on benchmarking models, we refer the reader to Section 1.3 of the SI.

\begin{table}[t]
\setlength{\tabcolsep}{5pt}
\renewcommand{\arraystretch}{1.2}
\caption{Recall and balanced accuracy score (BAS) for various classification methods.}
\begin{tabular}{l|ccc|c}
\hline
\hline
\multicolumn{1}{l|}{} & \multicolumn{3}{c|}{\textbf{Recall (\%)}} & \multicolumn{1}{l}{} \\
 & \textbf{HDA} & \textbf{LDA} & \textbf{MDA} & \textbf{BAS (\%)} \\ \hline
MLP & \textbf{98.5} & \textbf{77.8} & \textbf{80.6} & \textbf{85.6} \\
k-NN & 98.0 & 72.4 & 70.6 & 80.3 \\
Logistic Regression & 98.0 & 76.7 & 80.4 & 85.0 \\
Random Forest & 97.7 & 77.4 & 79.7 & 84.9 \\ \hline
\hline
\end{tabular}
\label{table:bechmarking_table}
\end{table}

\begin{figure}[t]
    \centering
    \includegraphics[width=\linewidth]{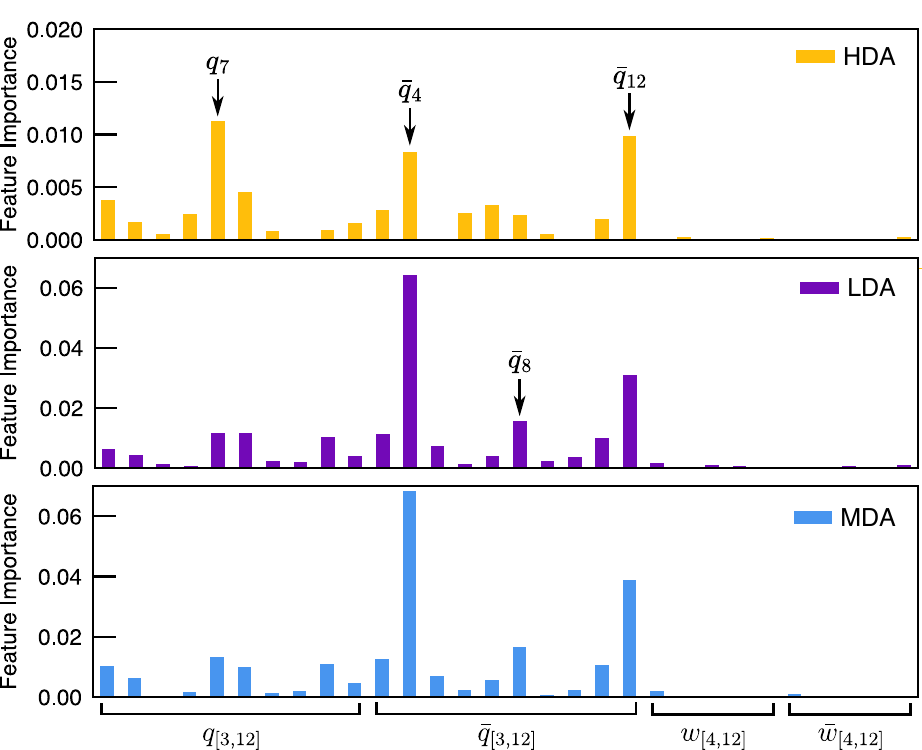}
    \caption{Permutation feature importance for the three NN models of amorphous ice phases.}
    \label{fig:sensitivity_analysis}
\end{figure}

To further probe the key characteristics of the different classes of amorphous ices, we performed sensitivity analysis on the input vectors (Fig.~\ref{fig:sensitivity_analysis}). To do, so we used permutation feature importance (PFI) as implemented in \texttt{scikit-learn} \cite{scikit-learn}. PFI gauges feature importance by measuring the reduction in a model's accuracy when a single feature value is randomly shuffled \cite{Breiman-01-10}. By disrupting the relationship between the feature and the target, this process reveals the extent to which the model relies on that specific feature. Given that each feature in the NN corresponds to a specific crystalline symmetry, this analysis allows us to determine the key symmetries which characterize each individual phase.

We built three distinct NN models using the optimized hyperparameters detailed above. The models were individually trained to distinguish between the presence or absence of each structure type (i.e., binary classification). Subsequently, we computed PFIs for each model. This approach allows us to discern the unique set of features and, consequently, the local symmetries that the model deems most important when classifying a local environment for each structure type.

\begin{figure*}[!ht]
    \centering
    \includegraphics[width=\textwidth]{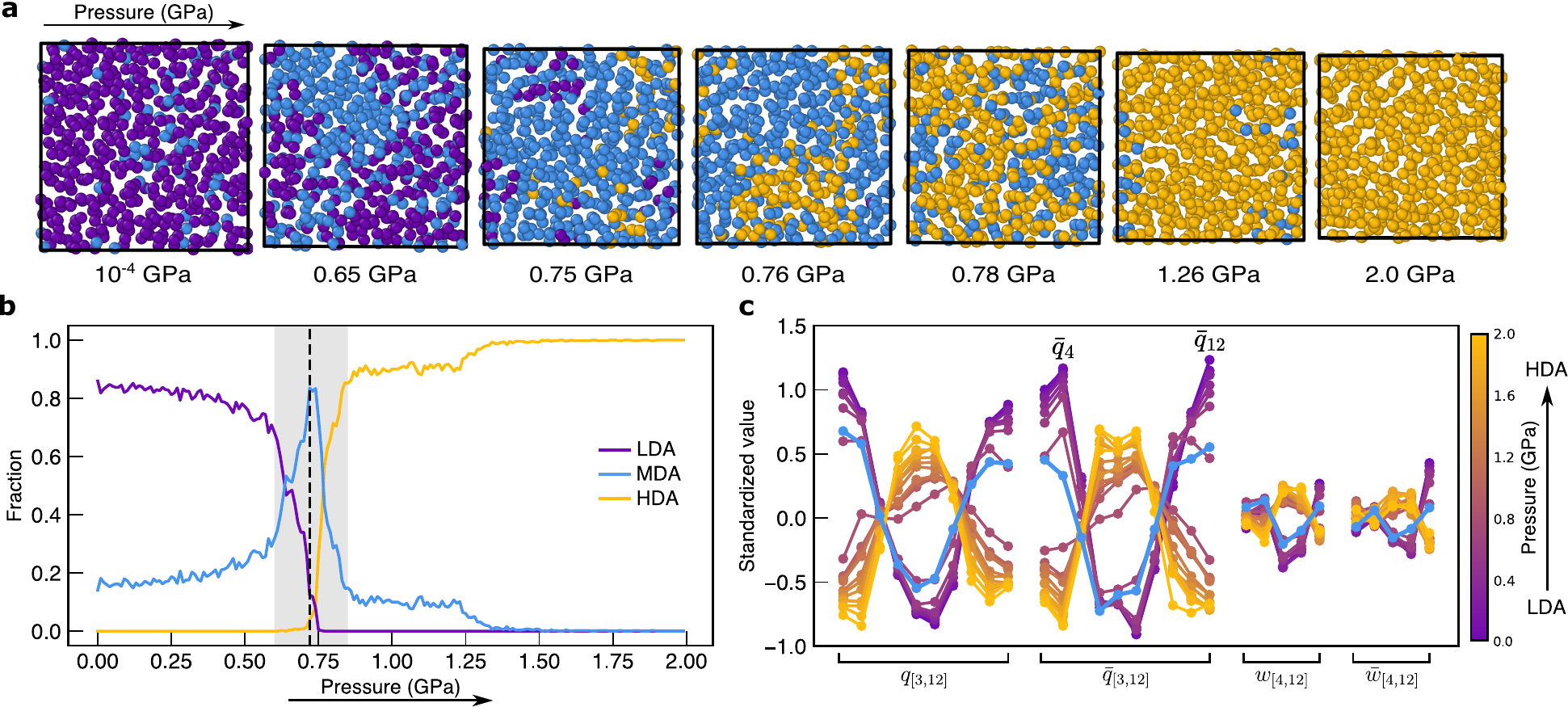}
    \caption{Classification analysis for the isothermal compression of LDA ice at $T=100$\,K. (a) Structure snapshots at various pressures along the compression trajectory from Ref.~\citenum{Martelli-20-9}. Atoms are color-coded according to which class of amorphous ice the NN model classified them as. Hydrogen atoms are omitted for clarity. Structures were visualized using OVITO \cite{Stukowski_2010}. (b) Fraction of local environments as classified by the NN as a function of pressure. The gray band shows the region of the phase transition. The dashed line represents the pressure at which the MDA structure most closely resembles a structure along the compression trajectory. (c) Evolution of the BOOs as a function of pressure.} 
    \label{fig:trajectories}
\end{figure*}
Figure \ref{fig:sensitivity_analysis} highlights the universal importance of $\bar{q}_{4}$ and $\bar{q}_{12}$ in characterizing the three amorphous ice phases. Beyond these, however, differences emerge. In particular, in the classification of HDA, the value of $q_7$ carries considerably more weight compared to its importance in LDA and MDA classification. Additionally, LDA and MDA show reduced dependency on features outside of $\bar{q}_{4}$ and $\bar{q}_{12}$ with perhaps a slight preference for $\bar{q}_{8}$ over other features. The strong similarity in feature importance between the LDA and MDA underscores again the high structural resemblance present in the local environments of both phases. This similarity serves as an explanation for the high misclassification ratios observed by the NN. Interestingly, the cubic order parameters show negligible influence in discerning the local environments across these phases. We also performed PFI calculations using Random Forest models (Section 2.2 of the SI).

To delve deeper into the nature and characteristics of MDA, and its relation to the other amorphous forms of water, we applied our NN to structures sampled from LDA compression trajectories taken from Ref.~\citenum{Martelli-20-9}. These trajectories were obtained via isothermal compression of LDA at three temperatures ($T=100$\,K,  $T=120$\,K, and $T=140$\,K), from $10^{-4}$ to $2.0$\,GPa, at a compression rate of 0.01 GPa/ns. Applying the order parameters to these compression structures, we use our trained NN to compute the fraction of each target state present within the structures. An illustrative set of structure snapshots along the trajectory is shown in Fig.~\ref{fig:trajectories}a, where each atom is color-coded according to the class of amorphous ice as which the environment was classified. The compositions of the different populations are shown in Fig.~\ref{fig:trajectories}b for $T=100$\,K. The graph shows expected behavior with LDA-like atoms dominating at low pressures. As pressure increases, this population declines to zero, coinciding with a sharp increase in the proportion of HDA-like atoms. This behavior is consistent with the known phase change from LDA to HDA~\cite{Martelli-18-7,Martelli-17-9}.

Interestingly, the proportion of MDA-like environments exhibits very different behavior. At lower pressures, there is a gradual and steady rise in the fraction of MDA-like environments present within the structures. During the phase change, a notable spike in the MDA-like population is observed (gray shaded area in Fig.~\ref{fig:trajectories}b), followed by a rapid decrease to zero once the transition to HDA is complete. This behavior suggests that MDA might occur as a transient state during the transition from LDA to HDA. 

In Fig.~\ref{fig:trajectories}c, we show the evolution of all BOOs considered herein as a function of pressure for $T=100$\,K. We standardize the BOOs across all atomic environments in the trajectory. This standardization highlights the behavior of all BOOs with pressure, revealing two distinct clusters: one at low pressures (purple) and another at high pressures (yellow), with a clear gap coincident with the phase transition. The plot distinctly showcases unique characteristics between LDA-like and HDA-like environments. These align with our knowledge of the short-range order in these two phases. The set of BOOs for each phase forms a structural signature which the NN can easily recognize. BOOs characterizing MDA-like environments, however, notably overlap with low-pressure BOOs resembling LDA. 

This plot also aligns with the outcomes of the sensitivity analysis (Fig.~\ref{fig:sensitivity_analysis}). Specifically, $\bar{q}_{4}$ and $\bar{q}_{12}$ maximally separate the LDA and HDA phases, enabling the NN to most easily use these features to distinguish between the two phases, thereby explaining the high PFI scores. In contrast, the $w$ and $\bar{w}$ parameters exhibit much greater noise and fewer defining characteristics resulting in negligible predictive power as shown by their extremely low PFI scores.

By comparing the trajectory BOOs with the MDA descriptor we can identify the structure and, consequently, the thermodynamic conditions along the LDA compression trajectory that best match the MDA structure. Our approach involves determining the minimum root-mean-square distance (RMSD) between the mean BOO from all MDA environments and the mean BOO of the environments at each pressure along the compression trajectory. We observe that the RMSE reaches its minimum during the phase transition (dashed line in Fig.~\ref{fig:trajectories}b), indicating that the closest resemblance to MDA local environments is found during the LDA to HDA shift. The same analysis for $T=120$\,K and $T=140$\,K is shown in Section 2.3 of the SI.

In conclusion, we combined high-dimensional order parameters and a NN model to probe the characteristics of amorphous ices. We showed that our optimized NN outperforms classical baseline classification models. Our analysis afforded insights into the nature of LDA, HDA, and the newly discovered medium-density amorphous ice (MDA). While previous work successfully categorized HDA, LDA, and liquid environments with minimal misclassification errors\cite{Martelli-20-9}, we showed that including MDA structures in the training dataset significantly reduces the model's performance. The difficulty of the problem lies in similarities in local structures between LDA and MDA that challenge clear delineation among the phases on structural grounds alone. Using sensitivity analysis on the high-dimensional OPs and our NN, we were able to extract key structural markers for the amorphous phases. Beyond water, we expect such approaches to be more general: for example, complex structural transitions under pressure are known for amorphous silicon \cite{Deb2001,Deringer2021}, and order parameters initially used for water \cite{Errington2001} have been applied to study tetrahedrality in phase-change memory materials \cite{Caravati-07-10}. Hence, our approach of combining BOOs and NN models could lead to a systematic framework for characterizing amorphous materials.

\begin{acknowledgments}
Z.F.B. thanks J.L.A. Gardner for useful discussions and advice. F.M. is grateful to Pablo Debenedetti for insightful discussions. Z.F.B. was supported through an Engineering and Physical Sciences Research Council DTP award and IBM Research. F.M. acknowledges support from the Hartree National Centre for Digital Innovation, a collaboration between STFC and IBM.
\end{acknowledgments}

\section*{Author Declaration}
\subsection*{Conflict of Interest}
The authors have no conflicts to disclose.
\subsection*{Author Contributions}
\textbf{Zoé Faure Beaulieu}: Conceptualization (equal); Formal analysis (lead); Investigation (equal); Methodology (equal); Software (lead); Writing - original draft (equal). \textbf{Volker Deringer}: Conceptualization (equal); Methodology (supporting); Funding acquisition (equal); Supervision (equal); Writing - original draft (equal). \textbf{Fausto Martelli}: Conceptualization (equal); Methodology (lead); Funding acquisition (equal); Supervision (equal); Writing - original draft (equal).

\section*{Data availability}

Data and code supporting the present study are openly available at \url{https://github.com/ZoeFaureBeaulieu/NN-amorphous-ices}. 

\bibliography{apssamp}

\end{document}


\maketitle
\pagebreak

\tableofcontents

\pagebreak

\section{Methods}
\subsection{\label{subsec: Steinhard OPs}Steinhardt Order Parameters}
In this work we use local bond orientational order parameters (OPs) which map the structural properties of the local environment surrounding each water molecule onto a high-dimensional vector. First introduced by Steinhardt et al. \cite{Steinhardt-83-7}, these OPs use spherical harmonics to encode the symmetries of the structures in the complex vector $q_{lm}(i)$ and its average over a defined set of neighbors $\bar{q}_{lm}(i)$ \cite{Lechner-08-9},

\begin{equation} \label{eq:1}
    \begin{split}
        &q_{lm}(i) = \frac{1}{|N_i|}\sum_{j\in N_i}Y_{lm}(\mathbf{r}_{ij}), \\
        &\bar{q}_{lm}(i) = \frac{1}{|N_i|+1}\sum_{k \in \{i,N_i\}}q_{lm}(k).
    \end{split}
\end{equation}

Here, $N_i$ is the neighborhood of of particle $i$ (this was defined as the $|N_i| = 16$ nearest neighbors in this work), $l$ and $m$ are integers with $m \in [-l,l]$, $Y_{lm}(r_{ij})$ are the spherical harmonics, and $r_{ij}$ is the position vector from particle $i$ to $j$. The set of $l$ spherical harmonics defines an orthonormal basis spanning the $(2l+1)$-dimensional representation of the rotation group SO(3) relating the irreducible representation of SO(3) and the symmetries of crystalline structure. As defined in Eq.\ref{eq:1} the OPs are dependent on the choice of reference frame, we therefore take the average over $m$ to obtain a rotationally invariant OP which encodes an intrinsic property of the structure. The rotationally invariant $q_l(i)$ and their averages $\bar{q}_{l}(i)$ are defined as:

\begin{equation}\label{eq:2}
    \begin{split}
        &q_{l}(i) = \sqrt{\frac{4\pi}{2l+1}\sum^{l}_{m=-l}|q_{lm}(i)|^2}, \\
        &\bar{q}_{l}(i) = \sqrt{\frac{4\pi}{2l+1}\sum^{l}_{m=-l}|\bar{q}_{lm}(i)|^2}
    \end{split}
\end{equation}

The cubic OPs $w_l(i)$ and their average $\bar{w}_l(i)$ are defined as
\begin{equation}\label{eq:3}
    \begin{split}
        &w_{l}(i)=\frac{\sum\limits_{m_1+m_2+m_3=0} 
        \left(\begin{smallmatrix}
        l & l & l\\
        m_1 & m_2 & m_3
        \end{smallmatrix}\right)
        q_{lm_1}(i)q_{lm_2}(i)q_{lm_3}(i)}{\left( \sum^{l}_{m=-l}|q_{lm}(i)|^{2} \right) ^{3/2}}, \\
        &\bar{w}_{l}(i)=\frac{\sum\limits_{m_1+m_2+m_3=0} 
        \left(\begin{smallmatrix}
        l & l & l\\
        m_1 & m_2 & m_3
        \end{smallmatrix}\right)
        \bar{q}_{lm_1}(i)\bar{q}_{lm_2}(i)\bar{q}_{lm_3}(i)}{\left( \sum^{l}_{m=-l}|\bar{q}_{lm}(i)|^{2} \right) ^{3/2}},
    \end{split}
\end{equation}
where the term in parentheses is the Wigner 3$j$ symbol.

\pagebreak

\subsection{\label{subsec: Database} Database}
The database was curated to contain a wide variety of amorphous ice structures. Where these originated from molecular dynamics (MD) simulations, water molecules were described by the TIP4P/2005 interaction potential \cite{Abascal-05-12}.

\subsubsection{LDA and HDA}
The bulk of the database was taken from work by Martelli et al. \cite{Martelli-20-9}. This consists of many LDA and HDA structures, each containing $8192$ water molecules. LDA structures were produced by simulating the quenching of equilibrated liquid water at $T=300$\,K to $T=80$\,K at a cooling rate of 1\,K/ns. HDA structures were produced by simulating the isothermal compression of LDA and ice I\textit{h} at four temperatures: $T=80$\,K,  $T=100$\,K,  $T=120$\,K, and $T=140$\,K. 
Isothermal decompression of HDA from 2.0\,GPa to $10^{-4}$\,GPa was then simulated to obtain configurations down to ambient pressure. 
For a detailed description of the simulation protocols, the reader is directed to Ref. \cite{Martelli-20-9}.

\subsubsection{MDA}
\begin{figure}[t]
    \centering
    \includegraphics[width=.5\linewidth]{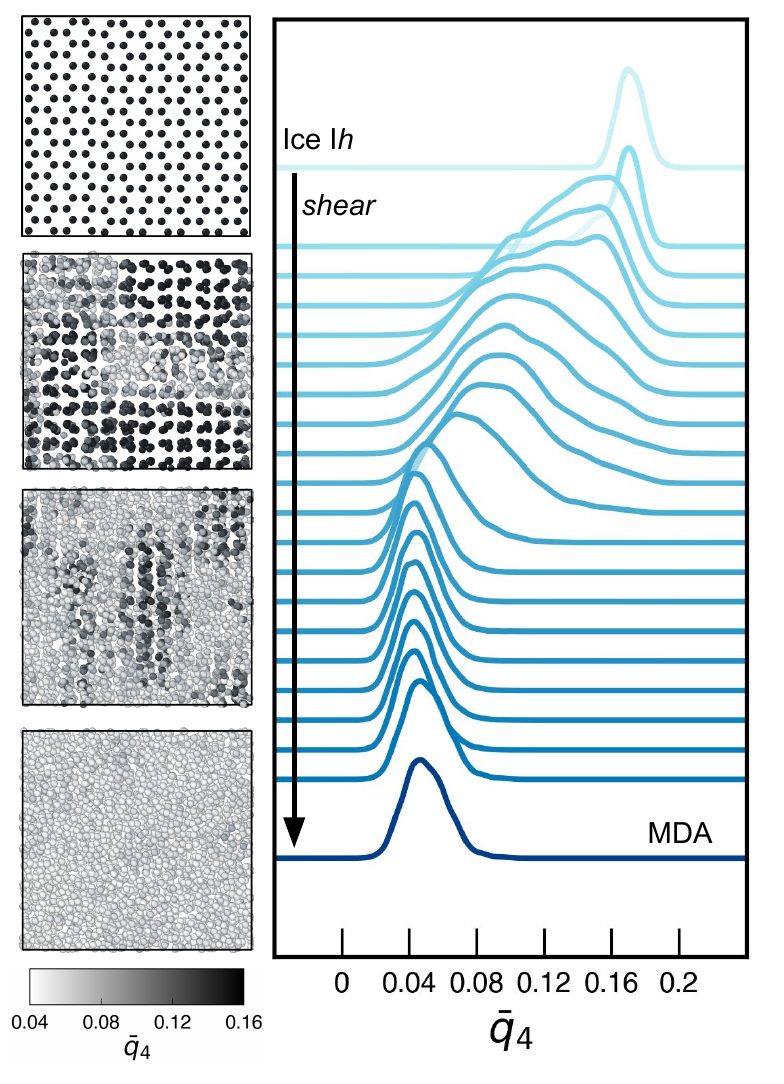}
    \caption{Steinhardt $\bar{q}_{4}$ values for I\textit{h} and their evolution during multiple shearing sequences to produce MDA. Left: Structure snapshots at various points during the shearing process. Atoms are colour-coded according to the value of $\bar{q}_{4}$. Right: Kernel density estimates (KDE) of Steinhardt $\bar{q}_{4}$ values in a 2880-molecule system at selected points along a shearing simulation (trajectory data from Ref. \cite{Rosu-Finsen-23-2}).}
    \label{fig:shear_snapshots}
\end{figure}

In addition to the above, a set of MDA configurations was taken from the work by \citep{Rosu-Finsen-23-2} and added to the database. 
MDA was produced by repeatedly shearing random layers of an ice I\textit{h} simulation box containing $2880$ water molecules. After each shear, geometry optimisation of the local molecular environments was performed. The shearing process was repeated until a number of structural characteristics converged, at which point the structure was deemed to have fully amorphised. For a $2880$ molecule box, structures were deemed fully amorphous after $100$ shears (Fig.\ref{fig:shear_snapshots}). Finally, the amorphised structures underwent an NPT MD simulation at 125\,K and 0\,atm with a time-step of 2\,fs. MDA structures were taken from MD production runs of 2\,ns. See Ref. \cite{Rosu-Finsen-23-2} for complete details. The final database includes the ice I\textit{h} structure as well as the final MDA structures and shear trajectories from 5 independent repeats.

\clearpage

\subsection{Benchmarking Classification methods}
\begin{figure}[h]
    \centering
    \includegraphics[width=\linewidth]{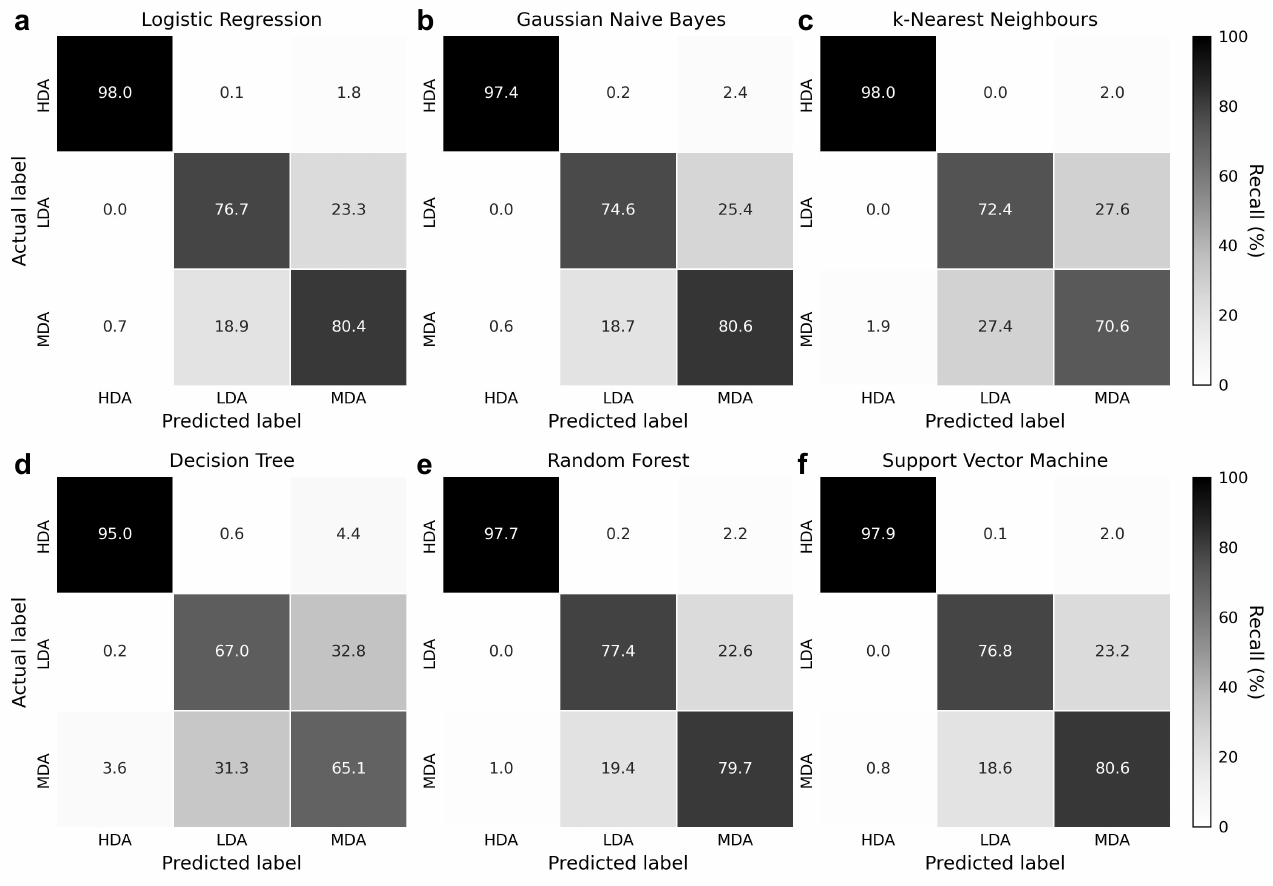}
    \caption{Confusion matrices for various ML classification algorithms. (a) Logistic Regression. (b) Gaussian Naive Bayes. (c) k-Nearest Neighbors with $k=5$. (d) Decision Tree. (e) Random Forest. (f) Support Vector Machine with an RBF kernel.}
    \label{fig:benchmarking_classification}
\end{figure}

We benchmarked our NN against 6 popular classification methods -- Logistic Regression, Naive Bayes, k-Nearest Neighbors (k-NN), Decision Trees, Random Forest, and Support Vector Machines (SVM) -- all as implemented in scikit-learn \cite{scikit-learn}. Fig. \ref{fig:benchmarking_classification} shows the confusion matrices for these methods. Our findings indicate that all 6 techniques perform similarly to our NN in classifying HDA environments, achieving a recall rate of over 95\%. However, the models differ in their ability to classify LDA and MDA environments. Logistic Regression, Naive Bayes, Random Forest and SVMs exhibit comparable performance to the NN, while k-NN and Decision Tree notably underperform in this regard.

The balanced accuracy scores \cite{5597285} for all the models investigated are reported in Table \ref{table:accuracy_scores}. We see that the NN model outperforms the spectrum of other classification algorithms analysed in this study.

\begin{table}[h]
\centering
\begin{tabular}{|c|c|c|c|c|c|c|}
\hline
NN & \begin{tabular}[c]{@{}c@{}}Logistic\\ Regression\end{tabular} & \begin{tabular}[c]{@{}c@{}}Gaussian\\ Naive Bayes\end{tabular} & k-NN & \begin{tabular}[c]{@{}c@{}}Decision \\ Tree\end{tabular} & \begin{tabular}[c]{@{}c@{}}Random\\ Forest\end{tabular} & SVM \\ \hline
\textbf{85.6} & 85.0 & 84.2 & 80.3 & 75.7 & 84.9 & 84.9 \\ \hline
\end{tabular}
\caption{Balanced accuracy scores (\%) for the NN and the 6 other classification methods investigated.}
\label{table:accuracy_scores}
\end{table}

\clearpage

\section{Results}

\subsection{Classification with HDA, LDA and high-T Liquid}
Our study builds on work done by Martelli et al. \cite{Martelli-20-9} in which a NN was used to partition high-dimensional space by training solely on HDA, LDA and high-T liquid structures. This section replicates and verifies the outcomes detailed in their research.

To accomplish this our training, validation and test sets were comprised of $32000$, $10000$, and $10000$ local oxygen environments respectively. These were randomly sampled in equal proportion from all 3 classes of structures: HDA, LDA, and high-T liquid configurations. Local environments were represented using 30-dimensional input vectors including all Steinhardt OPs as detailed in the main text. High-T liquid configurations were obtained using MD simulations taken at 300K and $10^{-4}$\,GPa. 

\begin{figure}[h]
    \centering
    \includegraphics[width=\linewidth]{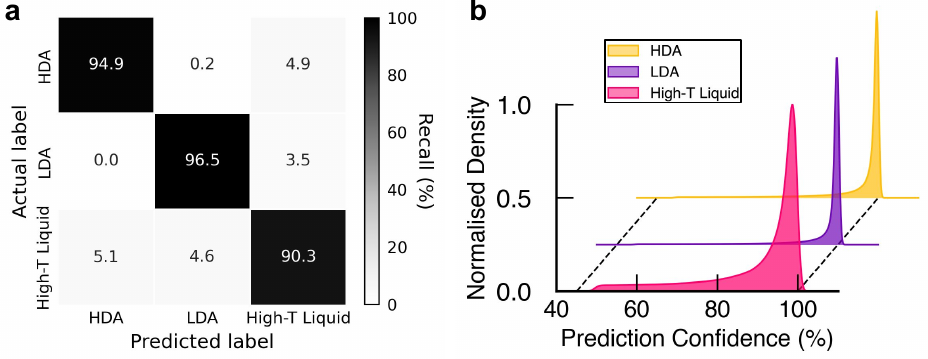}
    \caption{NN Classification with HDA, LDA and high-T liquid configurations. (a) Confusion matrix for local environment classification on the test set. (b) Prediction confidences for the three target states.}
    \label{fig:liquid_classification}
\end{figure}

We clearly see that the NN is able to achieve a misclassification ratio of $\sim5\%$ or less compared to $\sim18-23\%$ when high-T liquid is replaced with MDA. Similarly, the prediction confidences for the LDA phase are substantially higher when compared to Fig. 2(b) in the main text. These findings demonstrate the NN's ability to accurately categorize LDA environments based on Steinhardt descriptors. Conversely, the decrease in model performance when including MDA points towards these two phases having high resemblance in their local structures making differentiation by the model much more difficult. This resemblance is not present when only dealing with HDA, LDA and liquid phases.

\clearpage

\subsection{Sensitivity Analysis}
\begin{figure}[h]
    \centering
    \includegraphics[width=\linewidth]{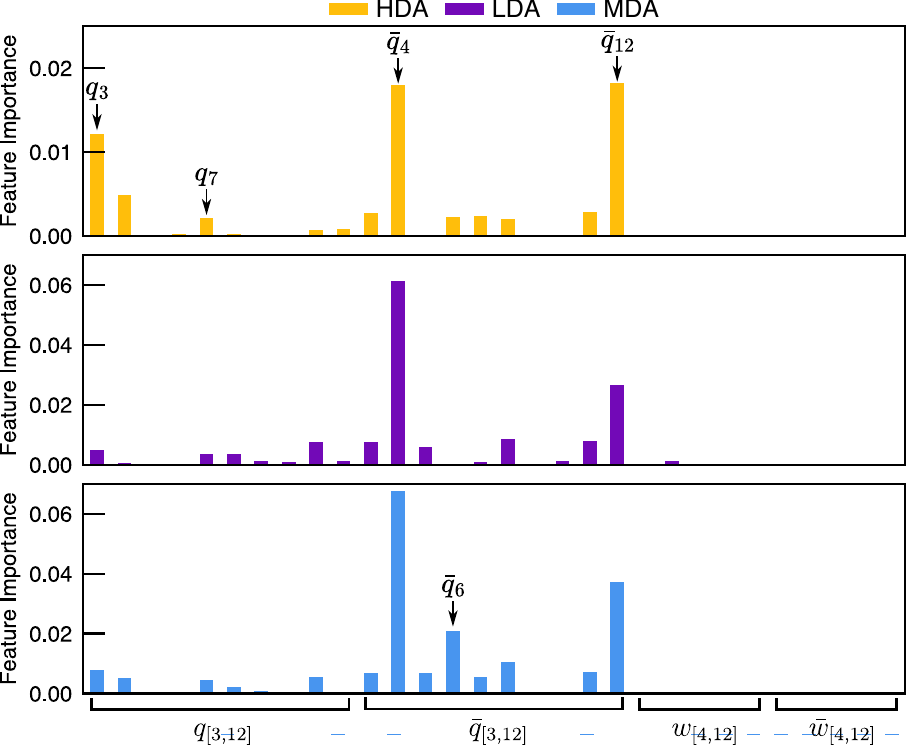}
    \caption{Permutation feature importance for the three Random Forest models of amorphous ice phases.}
    \label{fig:random_forest_pfi}
\end{figure}

We performed sensitivity analysis by calculating permutation feature importance (PFI) on three distinct NN models, each specialized in recognizing one of the three phases.  Additionally, we replicated this analysis using three Random Forest models instead. Fig. \ref{fig:random_forest_pfi} reveals that both the Random Forest and NN models highlight  $\bar{q}_{4}$ and $\bar{q}_{12}$ as being most important in characterizing the three phases. Both models also show no dependence on the cubic order parameters. However, there are some noteworthy disparities between the two model types.

Firstly, HDA shows a much stronger dependence on $q_3$ and places a much weaker emphasis on $q_7$ in the Random Forest model compared to the NN model. Secondly, a distinctive contrast surfaced between LDA and MDA regarding their reliance on $\bar{q}_{6}$. While the model disregards the $\bar{q}_{6}$ parameter entirely for LDA classification, it ranked it as the third most crucial feature when discerning MDA environments. This dissimilarity in behavior is absent when computing PFI values using a NN model (see Fig.\,3). 
These variations highlight the interpretational nuances between Random Forest and NN models in assessing feature importance and emphasize the difficult nature of the task at hand.

\clearpage

\subsection{LDA Compression Trajectories}
In this section we apply our NN to structures sampled from the LDA compression trajectories taken from Ref. \cite{Martelli-20-9}. These trajectories were taken at $T=120$\,K and $T=140$\,K. We use the previously trained NN to compute the fraction of each target state (HDA, LDA and MDA) present within the structures as shown for $T=100$\,K in Fig. 4.

\begin{figure}[h]
    \centering
    \includegraphics[width=\linewidth]{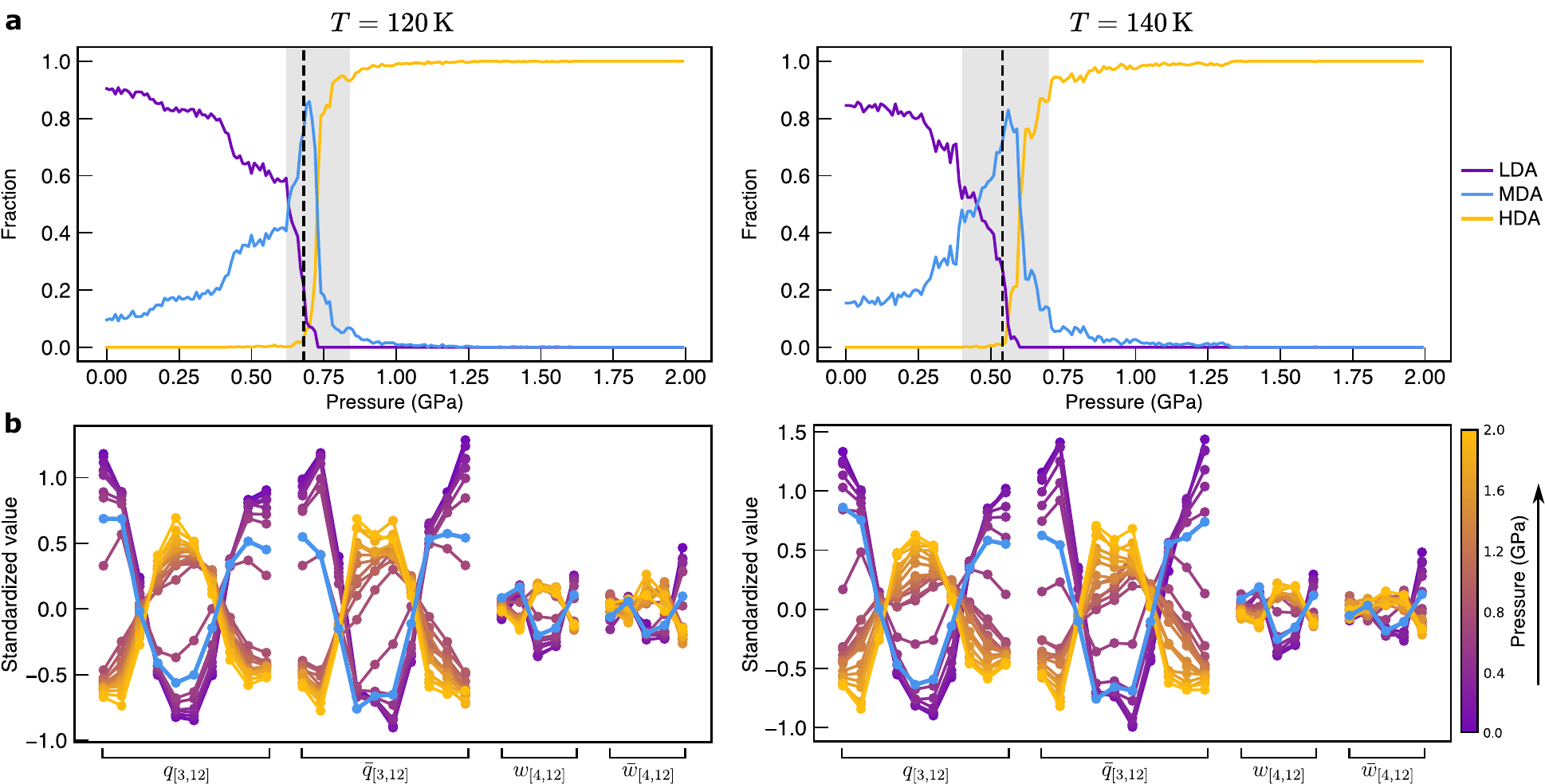}
    \caption{Classification analysis for the isothermal compression of LDA  at $T=120$\,K and $T=140$\,K. (a) Fraction of local environments as classified by the NN as a function of pressure. The grey band shows the region of the phase transition. The dashed line represents the pressure at which the MDA structure most closely resembles a structure along the compression trajectory. (c) Evolution of the Steinhardt parameters as a function of pressure.}
    \label{fig:trajectories_SI}
\end{figure}

Figure \ref{fig:trajectories_SI} shows that the behavior of the three phases at $T=120$\,K and $T=140$\,K is almost identical to that shown in Fig. 4.

\clearpage

\printbibliography